\documentstyle[12pt,epsf]{article}
\newcommand{\gsim}{\stackrel{>}{_\sim}}
\newcommand{\pk}{p_{_K}}
\begin{document}

\thispagestyle{empty}
\begin{flushright}
CERN-TH/97-374\\
LNF-98/003(P)\\
January 1998

\end{flushright}
\vspace*{1.5cm}
\centerline{\Large\bf Nonperturbative Effects in $\bar B\to X_sl^+l^-$}
\centerline{\Large\bf for Large Dilepton Invariant Mass}
\vspace*{2cm}
\centerline{{\sc Gerhard Buchalla${}^a$} and {\sc Gino Isidori${}^b$}}
\bigskip
\centerline{\sl ${}^a$Theory Division, CERN, CH-1211 Geneva 23,
                Switzerland}
\centerline{\sl ${}^b$INFN, Laboratori Nazionali di Frascati, 
                I-00044 Frascati, Italy}

\vspace*{1.5cm}
\centerline{\bf Abstract}
\vspace*{0.3cm}
\noindent 
We reconsider the calculation of ${\cal O}(\Lambda^2_{QCD}/m^2_b)$
nonperturbative corrections to $\bar B\to X_sl^+l^-$ decay.
Our analysis confirms the results of Ali et al. for the dilepton
invariant mass spectrum, which were in disagreement with an earlier
publication, and for the lepton forward-backward asymmetry. We also
give expressions for the ${\cal O}(\Lambda^2_{QCD}/m^2_b)$ corrections
to the left-right asymmetry. In addition we discuss the breakdown of
the heavy quark expansion near the point of maximal dilepton invariant
mass $q^2$ and consider a model independent approach to this region
using heavy hadron chiral perturbation theory. 
The modes $\bar B\to\bar Kl^+l^-$  and $\bar B\to\bar K\pi l^+l^-$,
which determine the endpoint region of the inclusive decay, 
are analyzed within this framework. 
An interpolation is suggested between the region of moderately high $q^2$, 
where the heavy quark expansion is still valid, and the vicinity of the
endpoint described by chiral perturbation theory.
We also comment on further nonperturbative
effects in $\bar B\to X_sl^+l^-$.
\vfill

\newpage
\pagenumbering{arabic}

\section{Introduction}

The inclusive decay $\bar B\to X_sl^+l^-$ ($l=e$, $\mu$, $\tau$) 
has received considerable interest in the literature \cite{HWS}--\cite{BURD}.
As a loop-induced flavour-changing neutral current (FCNC) process
it provides a sensitive probe of flavour dynamics, the least tested
sector of the Standard Model. The rare decay modes $\bar B\to X_sl^+l^-$
are well within reach of the next generation of precision $B$ physics
experiments and promise to yield much needed information complementary
to that from other sources such as $\bar B\to X_s\gamma$, 
$\bar B\to X_s\nu\bar\nu$,
$B-\bar B$ mixing, CP violation or rare $K$ decays. The interest in
$\bar B\to X_sl^+l^-$ and other inclusive rare $B$ decay processes is
reinforced by the fact that their theoretical treatment is fairly
well under control. Indeed, the rate for $\bar B\to X_sl^+l^-$ is dominated,
in the region of $q^2=(p_{l^-}+p_{l^+})^2$ away from resonance
backgrounds, by perturbatively calculable contributions. These are
known at next-to-leading order (NLO) \cite{MIS,BM}. Such a
calculation at the parton level is formally justified by the
heavy quark expansion (HQE), in which the free $b$ quark decay
$b\to sl^+l^-$ emerges as the leading contribution to 
$\bar B\to X_sl^+l^-$. This result receives power corrections of the form
$(\Lambda_{QCD}/m_b)^n$, which can be systematically addressed within the
HQE framework.
The leading corrections arise at order $n=2$. They have been first 
considered in \cite{FLS}. A further computation of these effects in
\cite{AHHM} did not confirm the results obtained in \cite{FLS}. 
In particular, in \cite{AHHM} the relative 
${\cal O}(\Lambda^2_{QCD}/m^2_b)$ correction diverges at the high-$q^2$
endpoint, indicating a manifest breakdown of the heavy quark expansion,
a feature that is absent in \cite{FLS}.
\\
The ${\cal O}(\Lambda^2_{QCD}/m^2_b)$ effects are relevant both conceptually,
for assessing the validity of the HQE, as well as for obtaining 
quantitative control over a class of theoretical uncertainties beyond
perturbation theory. In view of this, the phenomenological interest of
$\bar B\to X_sl^+l^-$ and the situation in the literature described above,
a further independent analysis of the issue is certainly useful.
The results of such an analysis of the ${\cal O}(\Lambda^2_{QCD}/m^2_b)$
corrections in $\bar B\to X_sl^+l^-$ will be presented in this
paper. 
We will furthermore discuss the breakdown of the HQE near the
endpoint of the spectrum (at maximum $q^2$). The implications of
this feature for a description of the high-$q^2$ region will be
pointed out. A major part of this work will then be devoted to
investigating the model independent constraints on the $q^2$-spectrum
that can be obtained using heavy hadron chiral perturbation theory
(HHChPT).

The paper is organized as follows. A brief description of the
general framework and a collection of basic formulas is given in
Section 2. In Section 3 we present our results for the $1/m^2_b$
corrections to the dilepton invariant mass spectrum, the 
forward-backward (FB) asymmetry and the left-right (LR) asymmetry
in $\bar B\to X_sl^+l^-$. Section 4 contains a discussion of the
breakdown of the HQE near the endpoint. In this section we also
analyze the endpoint region of $\bar B\to X_sl^+l^-$ in terms of
the exclusive modes $\bar B\to \bar Kl^+l^-$ and
$\bar B\to \bar K\pi l^+l^-$, calculated within chiral perturbation
theory. A few comments on further nonperturbative effects in
$\bar B\to X_sl^+l^-$ are made in Section 5. We summarize our
results in Section 6.

\section{Framework and Basic Expressions}

The starting point for the analysis of $\bar B\to X_sl^+l^-$ is the 
effective Hamiltonian, in the Standard Model given by
(neglecting the small contribution $\sim V^*_{us}V_{ub}$)
\begin{equation}\label{heff}
{\cal H}_{eff}=-\frac{G_F}{\sqrt{2}}V^*_{ts}V_{tb}
\left[\sum^8_{i=1}C_i(\mu) Q_i +
\frac{\alpha}{2\pi}\tilde C_9(\mu)(\bar sb)_{V-A}(\bar ll)_V +
\frac{\alpha}{2\pi}\tilde C_{10}(\bar sb)_{V-A}(\bar ll)_A \right]~.
\end{equation}
The Hamiltonian is known at next-to-leading order \cite{MIS,BM}.
A detailed review may be found in \cite{BBL}, where the Wilson
coefficients $C_i$ and the four-quark operators $Q_i$ are
defined explicitly (the operators are typically of the form 
$Q_i\sim(\bar sb)(\bar cc),$ for $i=1,\ldots, 6$, whereas
$Q_7\sim e m_b\bar s\sigma^{\mu\nu}(1+\gamma_5)b F_{\mu\nu}$ and 
$Q_8\sim g m_b\bar s\sigma^{\mu\nu}(1+\gamma_5)\lambda^a b G^a_{\mu\nu}$). 
\\
{}From (\ref{heff}) the following general expression can be derived
for the differential decay rate
\begin{eqnarray}\label{dg1}
&& \frac{d\Gamma(\bar B\to X_sl^+l^-)}{dx\ dy\ ds} = 
\frac{G^2_F m^5_b}{192\pi^3}
|V^*_{ts}V_{tb}|^2\frac{\alpha^2}{4\pi^2}\frac{3}{4\pi m^2_b}
\frac{m_b}{M_B}\times \\ && \ \ \times 
\left[L^S_{\mu\nu}\left\{\left(|\tilde C^{eff}_9|^2+|\tilde C_{10}|^2
\right)\ W^{\mu\nu}_9+4 m^2_b|C_7|^2\ W^{\mu\nu}_7+4 m_b {\rm Re}\ C_7
\tilde C^{eff*}_9\  W^{\mu\nu}_{97}\right\}  \right. \nonumber \\
&&\ \ \ \left. + L^A_{\mu\nu}\left\{2{\rm Re}\ \tilde C^{eff*}_9 \tilde C_{10}
\ W^{\mu\nu}_9+4 m_b {\rm Re}\ C_7\tilde C^*_{10}\ W^{\mu\nu}_{97}\right\}
\right]~. \nonumber
\end{eqnarray}
Here $m_b$ ($M_B$) is the $b$-quark ($B$ meson) mass.
$\tilde C^{eff}_9$ is a (scheme invariant) effective Wilson coefficient
that includes, in addition to $\tilde C_9$ from (\ref{heff}), the
contributions from the $b\to sl^+l^-$ transition matrix elements of
4-quark operators $Q_1$, $\ldots$, $Q_6$. Next
\begin{equation}\label{lsa}
L^S_{\mu\nu}=p_{1\mu}p_{2\nu}+p_{2\mu}p_{1\nu}-g_{\mu\nu} p_1\cdot p_2
\qquad \mbox{and} \qquad L^A_{\mu\nu}=
  -i\varepsilon_{\mu\nu\varrho\sigma}p^\varrho_1 p^\sigma_2
\end{equation}
are the symmetric and antisymmetric leptonic tensors,
respectively ($p_1$ ($p_2$) is the momentum of $l^-$ ($l^+$) and
$\varepsilon^{0123}=+1$). We also set
$s=q^2/m^2_b$ ($q=p_1+p_2$), $x=2 p\cdot p_1/m^2_b$ and
$y=2 p\cdot p_2/m^2_b$, where $p$  
is the $b$-quark momentum defined as $p^\mu=m_b v^\mu$
in terms of the $B$-meson four-velocity $v^\mu = p^\mu_B/M_B$. 
The hadronic tensors $W^{\mu\nu}_i$ can be written as 
$W^{\mu\nu}_i=2{\rm Im}\ T^{\mu\nu}_i$ where
\begin{eqnarray}\label{t97}
T^{\mu\nu}_9 &=& i\ \int d^4x\ e^{-iq\cdot x}
 \langle B|T\ j^{\dagger\mu}_9(x) j^\nu_9(0)|B\rangle~, \\
T^{\mu\nu}_{97} &=& i\ \int d^4x\ e^{-iq\cdot x}
 \langle B|T\ j^{\dagger\mu}_9(x) j^{\lambda\nu}_7(0)|B\rangle
 \frac{i q_\lambda}{q^2}~, \\
T^{\mu\nu}_{7} &=& i\ \int d^4x\ e^{-iq\cdot x}
 \langle B|T\ j^{\dagger\lambda\mu}_7(x) j^{\varrho\nu}_7(0)|B\rangle
 \frac{q_\lambda q_\varrho}{q^4}~, 
\end{eqnarray}
\begin{equation}\label{j97}
j^\mu_9=\bar s\gamma^\mu(1-\gamma_5)b~,\qquad\quad
j^{\mu\nu}_7=\bar s\sigma^{\mu\nu}(1+\gamma_5)b~.
\end{equation}
Here the $B$ meson state $|B\rangle$ is taken in conventional 
relativistic normalization $\langle B|B\rangle=2 E V$ (the explicit
appearance of the factor $1/M_B$ in (\ref{dg1}) is due to this
definition).
\\
Evaluating the hadronic tensors to leading order in the heavy quark
expansion, eq. (\ref{dg1}) reproduces the well known quark-level results
for the $\bar B\to X_sl^+l^-$ decay distributions and asymmetries.
For instance, defining
\begin{equation}\label{rsdef}
R(s)=\frac{\frac{d}{ds}\Gamma(\bar B\to X_sl^+l^-)}{
\Gamma(\bar B\to X_ce\nu)}~,
\end{equation}
one obtains upon integrating over $x$ and $y$
\begin{eqnarray}\label{rsnlo}
R(s) &=& \frac{\alpha^2}{4\pi^2}
\left|\frac{V_{ts}}{V_{cb}}\right|^2
\frac{(1-s)^2}{f(z)\kappa(z)}\times \\
&&\times\left[
(1+2s)\left(|\tilde C^{eff}_9|^2+|\tilde C_{10}|^2\right)+4
\left(1+\frac{2}{s}\right) |C_7|^2 + 12 C_7 \mbox{Re} \tilde C^{eff}_9
\right]~. \nonumber
\end{eqnarray}
Here $f(z)=1-8z^2+8z^6-z^8-24z^4\ln z$ 
is the phase space factor and $\kappa(z)$ the QCD correction
factor ($z=m_c/m_b$) entering $\Gamma(\bar B\to X_ce\nu)$; 
$\kappa(z)$ can be found in \cite{BBL}. 
Note that for the dilepton invariant mass spectrum $R(s)$ only the
symmetric part in (\ref{dg1}) (proportional to $L^S_{\mu\nu}$)
contributes. In (\ref{rsnlo}) we have neglected
${\cal O}(m^2_l/m^2_b)$ and  ${\cal O}(m^2_s/m^2_b)$
terms, as we shall do throughout this paper, unless stated otherwise. 
The expressions given are therefore applicable to the cases $l=e$, $\mu$.
The extensions of (\ref{rsnlo}) to the case $m_l\not=0$ 
(relevant for $l=\tau$) and $m_s\not=0$ are given in \cite{JoAnne,KS}.
Neglecting the strange quark mass is a very good approximation
except near the $q^2$ endpoint. This region, however, suffers 
from large nonperturbative corrections and the entire partonic
approach has to be reconsidered there (we will come back 
later to this point).

A quantity closely related to $R(s)$ is the left-right (LR) asymmetry,
which measures the difference in the rates of producing left handed
or right handed leptons in $\bar B\to X_sl^+l^-$ decay. As discussed
in \cite{BP,LD}, the LR asymmetry can be directly extracted
from (\ref{rsnlo}). Defining 
\begin{equation}
R^{L,R}(s) = R(s)\left |_{ 
\tilde C^{eff}_9\to\frac{\tilde C^{eff}_9\mp\tilde C_{10}}{2},\  
\tilde C_{10} \to \frac{\tilde  C_{10} \mp\tilde C^{eff}_9  }{2},\
|C_7|^2\to\frac{1}{2}|C_7|^2
 } \right. \\
\end{equation}
one has
\begin{eqnarray}\label{alr}
A_{LR}(s) &\equiv& R^L(s)-R^R(s)  \\
&= & \frac{\alpha^2}{4\pi^2}\left|\frac{V_{ts}}{V_{cb}}\right|^2
 \frac{(1-s)^2}{f(z)\kappa(z)}
 \left[(1+2s)\left(-2\tilde C_{10}\ {\rm Re}\ \tilde C^{eff}_9\right)
  -12 C_7 \tilde C_{10}\right]~. \nonumber
\end{eqnarray}
Another interesting observable that can be studied in 
$\bar B\to X_sl^+l^-$ decays is the forward-backward (FB) 
lepton asymmetry \cite{AMM}, which can be defined as 
\begin{equation}\label{asdef}
A_{FB}(s)=\frac{1}{\Gamma(\bar B\to X_ce\nu)}
  \int_{-1}^1 d\cos\theta ~
 \frac{d^2 \Gamma(\bar B\to X_s l^+l^-)}{d s~ d\cos\theta}
\mbox{sgn}(\cos\theta)~,
\end{equation}
where $\theta$ is the angle between $l^+$ and $B$ momenta 
in the dilepton center--of--mass frame. As shown in \cite{AHHM} 
$A_{FB}(s)$ is identical to the energy asymmetry introduced in \cite{CMW}. 
The NLO perturbative result for $A_{FB}(s)$ is given by
\begin{equation}\label{asnlo}
A_{FB}(s)= - \frac{3\alpha^2}{4\pi^2} \left|\frac{V_{ts}}{V_{cb}}\right|^2
\frac{ (1-s)^2}{f(z)\kappa(z)} 
\mbox{Re} \left\{\tilde C^*_{10}
\left[ 2\ C_7 + s\ \tilde C^{eff}_9 \right] \right\}~.
\end{equation}
Interestingly enough, both 
$A_{LR}(s)$ and $A_{FB}(s)$ are sensitive to the 
relative signs between  $C_{7}$,  $\tilde C^{eff}_9$
and  $\tilde C_{10}$. These asymmetries therefore offer 
useful additional information on the underlying short distance physics.

\section{${\cal O}(\Lambda^2_{QCD}/m^2_b)$ Power Corrections to
$R$, $A_{LR}$ and $A_{FB}$}

The hadronic tensors $W^{\mu\nu}_i$ in (\ref{dg1}) can be systematically
expanded in inverse powers of the heavy quark mass using the operator 
product expansion (HQE) approach supplemented by heavy quark effective
theory (HQET). The general procedure is described in great detail in
\cite{MW} for the case of $\bar B\to X_{u,c}l\nu$ decay. The first 
corrections to the parton result (${\cal O}(1)$) appear at 
${\cal O}(\Lambda^2_{QCD}/m^2_b)$. To this order we obtain the
following expressions for the hadronic tensors (after contracting
with $L^S_{\mu\nu}$)
\begin{eqnarray}
\frac{3}{4\pi m_b M_B}\int dx dy\ L^S_{\mu\nu}W^{\mu\nu}_{9} &=&
\left(1+\frac{\lambda_1}{2m^2_b}\right)(1-s)^2(1+2s) \nonumber\\
&&+\frac{3\lambda_2}{2m^2_b}(1-15s^2+10s^3)~,\quad \label{lw9}\\
\frac{1}{4\pi M_B}\int dx dy\ L^S_{\mu\nu}W^{\mu\nu}_{97} &=&
\left(1+\frac{\lambda_1}{2m^2_b}\right)(1-s)^2 \nonumber\\
&&-\frac{\lambda_2}{2m^2_b}(5+6s-7s^2)~, \label{lw97}\\
\frac{3m_b}{4\pi M_B}\int dx dy\ L^S_{\mu\nu}W^{\mu\nu}_{7} &=&
\left(1+\frac{\lambda_1}{2m^2_b}\right)(1-s)^2\left(1+\frac{2}{s}\right)
 \nonumber\\
&&-\frac{3\lambda_2}{2m^2_b}\frac{6+3s-5s^3}{s}~. \label{lw7}
\end{eqnarray}
Here
\begin{equation}\label{la12}
\lambda_1=\frac{\langle B|\bar h(iD)^2h|B\rangle}{2 M_B}~,\qquad
\lambda_2=\frac{1}{6}\frac{\langle B|\bar hg\sigma\cdot Gh|B\rangle}{2 M_B}
=\frac{M^2_{B^*}-M^2_B}{4}~,
\end{equation}
with $h$ the $b$-quark field in HQET.

The results in (\ref{lw9})--(\ref{lw7}) agree with \cite{AHHM} but
differ from the findings of \cite{FLS}. The contribution involving
$W^{\mu\nu}_9$ is the same that appears in the case of semileptonic
$\bar B\to X_ul\nu$ decay. Integration of (\ref{lw9}) over $s$ yields
$(1/2)[1+(\lambda_1-9\lambda_2)/(2m^2_b)]$, reproducing the well known
correction factor derived in \cite{MW,BIG}.
Inserting (\ref{lw9})--(\ref{lw7}) into (\ref{dg1}) we obtain for the
$1/m^2_b$ corrections to $R(s)$ in (\ref{rsnlo})
\begin{eqnarray}
\delta_{1/m^2_b}R(s)=\frac{3\lambda_2}{2m^2_b}\Biggl(
\frac{\alpha^2}{4\pi^2}\left|\frac{V_{ts}}{V_{cb}}\right|^2
\frac{1}{f(z)\kappa(z)}\Biggl[(1-15s^2+10s^3)(|\tilde C^{eff}_9|^2+
|\tilde C_{10}|^2) \nonumber 
\end{eqnarray}
\begin{equation}\label{dbr}
\qquad -(6+3s-5s^3)\frac{4|C_7|^2}{s}-(5+6s-7s^2)
4 C_7 \mbox{Re}\tilde C^{eff}_9\Biggr]+\frac{g(z)}{f(z)}R(s)\Biggr)~.
\end{equation}
Here we have used the normalizing semileptonic rate including terms
of order $1/m^2_b$
\begin{equation}\label{gslmb}
\Gamma(\bar B\to X_c e\nu)=\frac{G^2_F m^5_b}{192\pi^3}|V_{cb}|^2
f(z)\kappa(z)\left[1+\frac{\lambda_1}{2m^2_b}-\frac{3\lambda_2}{2m^2_b}
  \frac{g(z)}{f(z)}\right]~,
\end{equation}
\begin{equation}\label{gz}
g(z)=3-8z^2+24z^4-24z^6+5z^8+24z^4\ln z~,
\end{equation}
that can be found for instance in \cite{MW}.
Note that the correction due to the kinetic energy of the $b$-quark
$\sim\lambda_1$ is given as a simple overall factor
$(1+\lambda_1/(2m^2_b))$ for both $\bar B\to X_c e\nu$ and 
$\bar B\to X_sl^+l^-$
and therefore drops out in the ratio $R(s)$. Since, in contrast to 
$\lambda_2$, the quantity $\lambda_1$ is not well known anyway, its
absence in (\ref{dbr}) is a welcome feature.
\\
Given the results in (\ref{lw9})--(\ref{lw7}) it is straightforward
to write down the $1/m^2_b$ correction for $A_{LR}(s)$ in (\ref{alr})
\begin{eqnarray}\label{dbalr}
\delta_{1/m^2_b}A_{LR}(s)&=&\frac{3\lambda_2}{2m^2_b}\Biggl(
\frac{\alpha^2}{4\pi^2}\left|\frac{V_{ts}}{V_{cb}}\right|^2
\frac{1}{f(z)\kappa(z)}\Biggl[(1-15s^2+10s^3)
(-2\tilde C_{10}\ {\rm Re}\tilde C^{eff}_9)\nonumber \\
&&\ +(5+6s-7s^2)4 C_7 \tilde C_{10}\Biggr]+\frac{g(z)}{f(z)}A_{LR}(s)\Biggr)~.
\end{eqnarray}
This correction has been discussed previously in \cite{BP}, however
based on the incorrect results of \cite{FLS}.
\\
To calculate the FB asymmetry it is necessary to contract 
$W^{\mu\nu}_{9}$ and $W^{\mu\nu}_{97}$  
with the asymmetric component of the leptonic tensor.
The relevant terms, expanded up to ${\cal O}(\Lambda^2_{QCD}/m^2_b)$,
are given by
\begin{eqnarray}
\frac{1}{2\pi m_b M_B}\int dx dy\ \mbox{sgn}(y-x) \
 L^A_{\mu\nu}W^{\mu\nu}_{9}  &=& s(1-s)^2+ \frac{\lambda_1}{6m^2_b} 
s(3+2s+3s^2) \nonumber \\ 
&&-\frac{\lambda_2}{2m^2_b}s(9+14s-15s^2)~,\quad \label{law9}\\
\frac{1}{2\pi M_B}\int dx dy\ \mbox{sgn}(y-x) \
 L^A_{\mu\nu}W^{\mu\nu}_{97} &=& (1-s)^2 + \frac{\lambda_1}{6m^2_b} 
(3+2s+3s^2) \nonumber \\ 
&&-\frac{\lambda_2}{2m^2_b}(7+10s-9s^2)~, \label{law97}
\end{eqnarray}
leading to 
\begin{eqnarray}
\delta_{1/m^2_b}A_{FB}(s) = \frac{3\lambda_2}{2m^2_b}\Biggl(
\frac{\alpha^2}{4\pi^2}\left|\frac{V_{ts}}{V_{cb}}\right|^2
\frac{1}{f(z)\kappa(z)} \mbox{Re} \Biggl\{  \tilde C_{10}^*
\Biggl[ s(9+14s-15s^2) \tilde C^{eff}_9 \nonumber 
\end{eqnarray}
\begin{equation}\label{dbafb}
\qquad +(7+10s-9s^2) 2 C_7 \Biggr] \Biggr\}
  +\frac{g(z)}{f(z)}A_{FB}(s)\Biggr)+
\frac{4\lambda_1}{3m^2_b}\frac{s}{(1-s)^2}
A_{FB}(s)~.
\end{equation}
Also in this case our finding is in agreement with \cite{AHHM}. 

\section{The High-$q^2$ Region}

\subsection{Generalities}

The size of the ${\cal O}(\Lambda^2_{QCD}/m^2_b)$ corrections in
(\ref{dbr}), (\ref{dbalr}) and 
(\ref{dbafb}) is quite moderate, at the level of several percent,
for values of $s$ below about $0.6$. On the other hand, when
$s$ approaches the endpoint ($s=1$), the corrections for
$R$, $A_{LR}$ and $A_{FB}$ tend towards a nonzero value,
while the leading term of these quantities vanishes as $(1-s)^2$. The
relative correction thus diverges in the limit $s\to 1$ and the rate
$R(s)$ becomes even negative for $s$ close enough to the endpoint.
Obviously, the HQE breaks down in the endpoint region. {}From the 
expressions given above one may recognize that in the case of
$R(s)$ and $A_{LR}$ this behaviour is 
exclusively related to the $\lambda_2$-term, whereas in the case 
of  $A_{FB}$ also the kinetic energy correction $\sim\lambda_1$ 
is not well behaved in the limit $s\to 1$. 
We remark that these features are not shared by the result given in
\cite{FLS} and have been first observed by the authors of \cite{AHHM}.
\\
In order to account for nonperturbative effects that elude the HQE
approach, \cite{AHHM} supplement the partonic calculation with a
Fermi-motion model to predict the $q^2$ spectrum and the shape of
the FB asymmetry in the entire physical region including the endpoint.
Although such an approach could be useful in principle, in
particular when employed in conjunction with experimental data
(used e.g. to fit model parameters), we will not perform such an
analysis here. Instead, we would like to discuss to what extent
model independent predictions can be made for the $\bar B\to X_sl^+l^-$
spectrum. We will thereby focus our attention on the high-$q^2$ region.
For this purpose we shall first discuss the nature of the breakdown
of HQE in slightly more detail. We follow here the general discussion
presented in \cite{BSUV,MN}.

The central element in the operator product expansion of the
tensors $T^{\mu\nu}_i$ in (\ref{t97}) is the $s$-quark propagator.
This propagator emerges in the evaluation of the time-ordered
products (\ref{t97}) and determines essential features of the $1/m_b$
expansion. Denoting $k=m_b v-q$, $D_\mu=\partial_\mu-igA_\mu$
and neglecting the $s$-quark mass, the $s$-quark propagator in a
gluon background field may be written as
\begin{equation}\label{sprop}
S_s(k)=\frac{\not\! k+i\not\!\! D}{k^2+2ik\cdot D-
  \not\!\! D\not\!\! D+i\varepsilon}~.
\end{equation}
Up to terms of order $\Lambda_{QCD}\equiv\Lambda$, $k$ is
the momentum of the final state hadronic system. In the usual case,
that is away  from singular kinematical points, one has
$k\sim m_b$, $k^2\sim m^2_b$ and consequently the hierarchy
$k^2\sim m^2_b$ $\gg$ $k\cdot D\sim m_b\Lambda$ $\gg$
$\not\!\! D\not\!\! D\sim\Lambda^2$.
Therefore one can expand $S_s(k)=\not\! k/k^2+{\cal O}(\Lambda/m_b)$
and the usual HQE is valid.
\\
A different situation arises in the endpoint region of the lepton energy
spectrum in $\bar B\to X_{c,u}l\nu$ and for the photon energy spectrum in
$\bar B\to X_s\gamma$. Here one still has $k\sim m_b$ in terms of components,
however the kinematics is now such that $k^2\sim m_b\Lambda$. For the
quantities in the denominator of (\ref{sprop}) this implies
$k^2\sim m_b\Lambda\approx k\cdot D\sim m_b\Lambda\gg
\not\!\! D\not\!\! D\sim\Lambda^2$.
An expansion in $\Lambda/m_b$ is still possible, but $k\cdot D/k^2$
is now of ${\cal O}(1)$ and the corresponding effects have to be
resummed to all orders. This is the case discussed in detail in
\cite{MN,NEU1} for $\bar B\to X_{c,u}l\nu$ and in \cite{NEU2} for
$\bar B\to X_s\gamma$ (see also \cite{BSUV}).
\\
We would like to stress that the situation encountered in the endpoint
region of the $q^2$ spectrum in $\bar B\to X_sl^+l^-$ is substantially
different from the two cases just described. For the kinematics that is
relevant here, $q^2\approx m^2_b\approx M^2_B$, it follows that
$k\sim\Lambda$ and $k^2\sim\Lambda^2$. Then all three terms in the
denominator of (\ref{sprop}) are of the same order of magnitude 
$\sim\Lambda^2$. The heavy quark expansion breaks down completely and
not even an all-orders resummation, of the type useful for
$\bar B\to X_{c,u}l\nu$ and $\bar B\to X_s\gamma$, can be performed. This
conclusion is clear on physical grounds, since at $q^2\approx M^2_B$
the two leptons are emerging back-to-back, carrying almost all the energy
released in the decay of the $B$ meson. The final state hadronic system
has very low momentum and we are in a regime of manifestly nonperturbative
QCD.

At this point we would like to emphasize a conceptual consequence of this
discussion for the treatment of the $q^2$ spectrum in $\bar B\to X_sl^+l^-$
within a Fermi-motion model, as employed in \cite{AHHM}. In the case of
the photon energy spectrum in $\bar B\to X_s\gamma$ or the lepton energy
spectrum in $\bar B\to X_{c,u}l\nu$ the resummation of leading singular
contributions in the HQE leads to a description of the endpoint
region in terms of a shape function \cite{NEU1,NEU2,BSUV}. The shape function
depends on nonperturbative physics that can qualitatively, at least to
some extent, be modeled by a Gaussian description of the $b$-quark
momentum distribution inside the $B$ meson (Fermi-motion). As explained
above, a similar interpretation does not exist in the case of 
$\bar B\to X_sl^+l^-$. A Fermi-motion description of nonperturbative effects,
particularly for high $q^2$, appears therefore certainly less justified
than in the usual applications to $\bar B\to X_s\gamma$ and 
$\bar B\to Xl\nu$.
In fact, as we have seen above, the divergence of the $1/m^2_b$
corrections to the $q^2$ spectrum
near the endpoint arises from the chromomagnetic
interaction term $\sim\lambda_2$ that does not have an obvious 
interpretation in terms of a Fermi-motion ansatz.

On the other hand, the kinematical situation in $\bar B\to X_sl^+l^-$
near the $q^2$ endpoint, with few, low-momentum hadrons in the
final state, lends itself to a treatment using heavy hadron chiral
perturbation theory (HHChPT) \cite{WI,BD}.
Combining this description at very high $q^2$ with the standard HQE
result at somewhat lower $q^2$, where the latter is still valid,
a model independent analysis of the entire high-$q^2$ region
(above the $\Psi$ and $\Psi'$ resonances) could be conceived.
In the following we shall examine such a possibility.

First, one may write down an effective Hamiltonian, suitable for the
endpoint region ($q^2\to M^2_B$) in $\bar B\to X_sl^+l^-$.
This Hamiltonian differs from the standard Hamiltonian (\ref{heff}).
`Light' quark ($u$, $d$, $s$, $c$) loops
may be integrated out explicitly since they involve the hard external
scale $q^2\sim m^2_b\gg 1 GeV$. This endpoint effective Hamiltonian then
takes the form, valid at NLO in QCD
\begin{eqnarray}\label{heffep}
&&{\cal H}_{eff,EP}=-\frac{G_F}{\sqrt{2}}V^*_{ts}V_{tb}
   \frac{\alpha}{2\pi} \times \\
&&\ \ \times\left[\tilde C_{9,EP}(\bar sb)_{V-A}(\bar ll)_V+
\tilde C_{10}(\bar sb)_{V-A}(\bar ll)_A+
2 m_b C_7\ \bar s\sigma^{\mu\nu}(1+\gamma_5)b\frac{iq_\mu}{q^2}
\bar l\gamma_\nu l\right]~. \nonumber
\end{eqnarray}
The Wilson coefficient $\tilde C_{9,EP}$ has the structure
\begin{equation}\label{c9ep}
\tilde C_{9,EP}=\tilde C^{NDR}_9+h(z,s)(3 C^{(0)}_1+C^{(0)}_2)+
\ (\mbox{penguin contributions})~,
\end{equation}
with $C^{(0)}_1$, $C^{(0)}_2$, $\tilde C^{NDR}_9$ from (\ref{heff}).
These quantities, the function $h(z,s)$ and the remaining terms can be
found in \cite{BBL}. $\tilde C_{9,EP}$ is identical to $\tilde C^{eff}_9$
in (\ref{dg1}) (see also \cite{BBL}), except that it does not include
the QCD correction $\tilde\eta(s)$ to the matrix element of the
current $(\bar sb)_{V-A}$, which multiplies $\tilde C^{NDR}_9$ in
$\tilde C^{eff}_9$.

The Hamiltonian (\ref{heffep}) is still normalized at a scale
$\mu={\cal O}(m_b)$. A further evolution down to hadronic scales
$\sim 1~\rm{GeV}$ is calculable perturbatively using HQET
(`hybrid renormalization'). However the HQET logarithms will be
automatically contained in the matrix elements of the 
$(\bar s\Gamma b)$ operators if they are taken in full QCD
and appropriately normalized at $\mu={\cal O}(m_b)$. It is therefore
not necessary to make these effects explicit in (\ref{heffep}).

\subsection{$\bar B\to\bar K l^+l^-$}

At the very high end of the spectrum, between the $K\pi$ threshold and
the physical endpoint, the inclusive decay $\bar B\to X_sl^+l^-$
degenerates into the exclusive mode $\bar B\to\bar Kl^+l^-$. Introducing the
variable $s_m\equiv q^2/M^2_B$, this region corresponds to
$s^{K\pi}_m \leq s_m \leq s^K_{m}$ where 
$s_m^K=s_{m,max}=(1-M_K/M_B)^2=0.821$ and $s^{K\pi}_m=0.774$.

The matrix elements needed for $\bar B\to\bar Kl^+l^-$ can be written as
\begin{equation}\label{fpmdef}
\langle\bar K(\pk)|\bar s\gamma_\mu(1-\gamma_5)b|\bar B(p)\rangle=
f_+(q^2)(p+\pk)_\mu+f_-(q^2)(p-\pk)_\mu~,
\end{equation}
\begin{equation}\label{atdef}
\langle\bar K(\pk)|\bar s\sigma^{\mu\nu}b|\bar B(p)\rangle=
-i a_T(q^2)(\pk^\mu p^\nu-\pk^\nu p^\mu)~,
\end{equation}
in terms of the form factors $f_\pm(q^2)$ and $a_T(q^2)$.
The decay rate (normalized to the semileptonic width 
as in (\ref{rsdef})) is then given by \cite{CDSS}
\begin{eqnarray}\label{rksm}
&&R_K(s_m) \equiv
\frac{\frac{d}{ds_m}B(\bar B\to\bar Kl^+l^-)}{B(\bar B\to X_ce\nu)}=
\frac{\tau(B_d)}{B(\bar B\to X_ce\nu)}\frac{G^2_FM^5_B}{192\pi^3}
|V_{tb}V_{ts}|^2\frac{\alpha^2}{4\pi^2}f_1(s_m)\times \nonumber\\
&&\ \ \times \left\{\frac{f^2_+}{2}\left(|\tilde C_{9,EP}|^2+|\tilde C_{10}|^2
\right)+\frac{a^2_T}{2}m^2_b|C_7|^2-f_+ a_T m_b {\rm Re}\
C_7\tilde C^*_{9,EP}\right\}~, 
\end{eqnarray}
where the phase space function $f_1$ reads
\begin{equation}\label{f1sm}
f_1(s_m)=((1-\varrho+s_m)^2-4s_m)^{3/2}~,\qquad\quad 
\varrho=\frac{M^2_K}{M^2_B}~.
\end{equation}
In general, the form factors are very difficult to calculate. However,
as long as we are interested in high $q^2$, where the kaon momentum
is small, HHChPT may be used to estimate these nonperturbative
quantities. In this approach, to the lowest order, 
one finds \cite{WI,BD,FG,CDDGFN}
\begin{equation}\label{fpmexp}
f_\pm=-\frac{f_B}{2f_\pi}\left(1\pm g
\frac{M_B\mp v\cdot \pk}{v\cdot \pk+\Delta+\mu_s}\right)~,\qquad\quad
v\cdot \pk=\frac{M^2_B+M^2_K-q^2}{2M_B}~,
\end{equation}
\begin{equation}\label{atexp}
a_T=\frac{g f_B}{f_\pi}\frac{1}{v\cdot\pk+\Delta+\mu_s}~.
\end{equation}
Here $f_B$ and $f_\pi$ are the $B$ meson and the pion decay constants
in the normalization where $f_\pi=132\,\mbox{MeV}$.
$\Delta=M_{B^*}-M_B=46\,\mbox{MeV}$,
$\mu_s=M_{B_s}-M_B=90\,\mbox{MeV}$ and  $g$ is the 
HHChPT parameter that  determines $B^*B\pi$ and $D^*D\pi$ 
couplings at low energy. The value of $g$ 
could in principle be inferred from a measurement of
$\Gamma(D^*\to D\pi)$, but present data
only allow to set an upper limit $g^{exp}<0.7$. 
According to the theoretical estimates of \cite{CDDGFN}
in the following we will assume $0.4<g<0.6$.

\begin{table}
\begin{center}
\begin{tabular}{|c|c|c|c|c|}\hline\hline
$m_b$ & $m_c$ & $|V_{cb}|$ & $|V_{ts}|$ & $|V_{tb}|$ \\
\hline
4.8 GeV & 1.4 GeV & 0.04 & 0.04 & 1 \\
\hline\hline
$\bar m_t(m_t)$ & $M_W$ & $\sin^2\Theta_W$ & $\alpha^{-1}$ & 
    $\Lambda^{(5)}_{\overline{MS}}$ \\
\hline
170 GeV & 80.2 GeV & 0.23 & 129 & 0.225 GeV \\
\hline\hline
$M_B$ & $M_K$ & $M_\pi$ & $\tau(B_d)$ & $B(\bar B\to X_ce\nu)$ \\
\hline
5.28 GeV & 0.496 GeV & 0.140 GeV & 1.6 ps & 0.104 \\
\hline\hline
$g$ & $f_B$ & $f_\pi$ & $\Delta=M_{B^*}-M_B$ & $\mu_s=M_{B_s}-M_B$ \\
\hline
0.5 & 0.180 GeV & 0.132 GeV & 0.046 GeV & 0.090 GeV \\
\hline\hline
\end{tabular}
\end{center}
\caption[]{Compilation of input parameters (central values).
\label{input}}
\end{table}

\subsection{$\bar B\to\bar K\pi l^+l^-$}

Between $K\pi$ and $K\pi\pi$ thresholds, 
i.e. for $s^{K\pi\pi}_m=0.728 \leq s_m \leq s^{K\pi}_{m}$,
also the $\bar B\to\bar K\pi l^+l^-$ decay is kinematically
allowed. No other modes are permitted and the hadronic invariant
mass is still small enough to justify the use of HHChPT.

The matrix element of the left-handed current 
relevant to  $\bar B\to\bar K\pi l^+l^-$ can be generally 
decomposed in terms of four independent form factors. Defining 
\begin{equation}\label{abchdef}
\langle\bar K^i(\pk)\pi^j(p_\pi)|\bar s\gamma_\mu(1-\gamma_5)b|
\bar B(p)\rangle=
i c_{ij} \left[ a p_{\pi,\mu} + b p_{K,\mu} + c p_\mu
       -2 i h \varepsilon_{\mu\alpha\beta\gamma}p^\alpha \pk^\beta
       p_\pi^\gamma \right] ~,
\end{equation}
the lowest order HHChPT results are given by \cite{LLW}
\begin{eqnarray}
  \label{abchth}
a &=& \frac{gf_B}{f_\pi^2} \frac{M_B}{v\cdot p_\pi+\Delta}~, 
\qquad\qquad b=0~,\\
c &=& \frac{f_B}{2f_\pi^2}\left[1
-2g\frac{v\cdot p_\pi}{v\cdot p_\pi+\Delta}
-\frac{v\cdot (\pk-p_\pi)}{v\cdot (\pk+p_\pi)+\mu_s}\right. \nonumber\\
&&\left.\ \ \ -2g^2 \frac{\pk\cdot p_\pi -v\cdot\pk v\cdot p_\pi}
{[v\cdot p_\pi+\Delta][v\cdot (\pk+p_\pi)+\mu_s]} \right],\\
\label{hchpt}
h &=& \frac{g^2 f_B}{2f_\pi^2} \frac{1}{ 
[v\cdot p_\pi+\Delta][v\cdot (\pk+p_\pi)+\Delta+\mu_s]}~,
\end{eqnarray}
where $|c_{-+}|^2= |c_{0-}|^2= 2|c_{00}|^2= 2|c_{-0}|^2=1$.

We have checked the results (\ref{abchth})--(\ref{hchpt}), 
first obtained by the
authors of \cite{LLW}, and agree with their findings. In addition
we need the corresponding matrix element of the magnetic penguin
operator. We obtain, again to leading order in HHChPT
\begin{eqnarray}\label{abchp}
&&\langle\bar K^i(\pk)\pi^j(p_\pi)|
\bar s\sigma_{\nu\mu}(1+\gamma_5)b\frac{i q^\nu}{q^2}|\bar B(p)\rangle=
\\
&&\ \ \ \ i c_{ij} \left[ a' p_{\pi,\mu} + b' p_{K,\mu} + c' p_\mu
       -2 i h' \varepsilon_{\mu\alpha\beta\gamma}p^\alpha \pk^\beta
       p_\pi^\gamma \right] ~, \nonumber
\end{eqnarray}
\begin{eqnarray}\label{abchpth}
a' &=& \frac{g f_B M_B}{f^2_\pi q^2(v\cdot p_\pi+\Delta)}
 \Biggl[M_B-v\cdot\pk-v\cdot p_\pi \nonumber\\
&&\ \ \ +g\frac{v\cdot\pk v\cdot(\pk+p_\pi)-
  \pk\cdot p_\pi-M^2_K}{v\cdot(\pk +p_\pi)+\Delta+\mu_s}\Biggr]~,\\
b' &=& \frac{g^2 f_B M_B}{f^2_\pi q^2(v\cdot p_\pi+\Delta)}
 \frac{\pk\cdot p_\pi+M^2_\pi-v\cdot p_\pi\ v\cdot(\pk+p_\pi)}{
       v\cdot(\pk +p_\pi)+\Delta+\mu_s}~,\\
c' &=& -\frac{g f_B}{f^2_\pi q^2(v\cdot p_\pi+\Delta)}
 \Biggl[M_B v\cdot p_\pi-M^2_\pi-\pk\cdot p_\pi \nonumber\\
&&\ \ \ +g\frac{\pk\cdot p_\pi\ v\cdot(\pk-p_\pi)-M^2_K v\cdot p_\pi+
  M^2_\pi v\cdot\pk}{v\cdot(\pk +p_\pi)+\Delta+\mu_s}\Biggr]~,\\
h' &=& \frac{g f_B}{2 f^2_\pi q^2(v\cdot p_\pi+\Delta)}
 \left[1+g\frac{M_B-v\cdot\pk- 
      v\cdot p_\pi}{v\cdot(\pk +p_\pi)+\Delta+\mu_s}\right]~.
\end{eqnarray}
We proceed to compute the decay rate.
The necessary four-body phase space integrations can be performed using
the general methods reviewed in \cite{LLW}. The leading behaviour of the
differential $\bar B\to\bar K\pi l^+l^-$ decay rate as a function of
$(s^{K\pi}_{m}-s_m)$ close to the $K\pi$ threshold may be written down
analytically. It gives the correct asymptotic behaviour at threshold
and can be used as an approximation to the full result for values of
$s_m$ not too far from this point. We find
\begin{eqnarray}
R_{K^-\pi^+}(s_m) \equiv 
\frac{\frac{d}{ds_m}B(\bar B\to K^-\pi^+l^+l^-)}{B(\bar B\to X_ce\nu)}=
\frac{\tau(B_d)}{B(\bar B\to X_ce\nu)}\frac{G^2_FM^5_B}{192\pi^3}
|V_{tb}V_{ts}|^2\frac{\alpha^2}{4\pi^2}\times \nonumber
\end{eqnarray}
\begin{equation}\label{rkpsm}
\ \ \times\frac{1}{32\pi^2}
\left\{F_9(s_m)\left(|\tilde C_{9,EP}|^2+|\tilde C_{10}|^2
\right)+4 F_7(s_m)|C_7|^2 + 4 F_{97}(s_m) {\rm Re}\
C_7\tilde C^*_{9,EP}\right\}~,
\end{equation}
\begin{equation}\label{f9sm}
F_9(s_m)=\frac{\pi}{4}\frac{(t_1 x_1 x_2)^{1/2}}{(1-\sqrt{t_1})^{3/2}}
 \left[ w^2_1+\frac{4 x_1 x_2}{t_1}(1-\sqrt{t_1})w^2_2\right]
  (s^{K\pi}_{m}-s_m)^3 ~,
\end{equation}
with $x_1=M_\pi/M_B$, $x_2=M_K/M_B$, $t_1=(x_1+x_2)^2$ and
$s^{K\pi}_{m}=(1-x_1-x_2)^2$. The functions $F_7$ and $F_{97}$ are
obtained from $F_9$ by replacing $w^2_i\to w'^2_i$ and
$w^2_i\to w_i w'_i$, respectively, where
\begin{equation}\label{w12}
w_1=\frac{f_B M_B}{f^2_\pi}\left[\frac{g M_\pi}{M_\pi+\Delta}
 \left(\frac{M_B}{M_K+M_\pi}-1\right)+
 \frac{M_\pi+\mu_s/2}{M_K+M_\pi+\mu_s}\right]~,
\end{equation}
\begin{equation}
w_2=-\frac{f_B M_B}{2 f^2_\pi}\frac{g M_B}{M_\pi+\Delta}~,
\end{equation}
\begin{equation}\label{w12p}
w'_1=\frac{f_B M_B}{f^2_\pi}\frac{g M_\pi}{M_\pi+\Delta}
 \frac{m_b}{M_K+M_\pi}~,\qquad\quad
w'_2=-\frac{f_B M_B}{2 f^2_\pi}\frac{g M_B}{M_\pi+\Delta}
 \frac{m_b}{M_B-M_K-M_\pi}~.
\end{equation}
Adding the two isospin channels, the total result for the 
nonresonant $\bar B_d(B^-)\to\bar K\pi l^+l^-$ rate becomes
\begin{equation}\label{rkptot}
R_{K\pi}\equiv R_{K^-\pi^+}+R_{\bar K^0\pi^0}=
 R_{\bar K^0\pi^-}+R_{K^-\pi^0}=\frac{3}{2} R_{K^-\pi^+}~.
\end{equation}
With the explicit expressions for $R_{K\pi}$ at hand, we are in
a position to estimate the relative importance of the nonresonant
$K\pi$ mode relative to the single $K$ channel in the endpoint region.
It is clear that the four-body process $\bar B\to\bar K\pi l^+l^-$
is phase space suppressed. This is obvious from (\ref{rkpsm}), which
exhibits the typical factor of $\sim 1/(16\pi^2)$. More
quantitatively we find that $R_{K\pi}$ amounts to less than $2\%$ of
$R_K$ at $s_m=0.7$ and is still less important for larger $s_m$.
$R_{K\pi}$ is therefore negligible in the entire endpoint region,
which is completely dominated by $R_K$.
The asymptotic formula (\ref{rkpsm}) describes the behaviour of
$R_{K\pi}$ close to threshold ($s_m=0.774$). For $s_m=0.7$
(\ref{rkpsm}) overestimates the full result by about 50\%. This
is still useful for an order of magnitude estimate. \\
For $s_m$ below 0.7 a substantial enhancement of the $K\pi$ mode 
is expected due to the contribution of the $K^*$ resonance. 
However, for $s_m>0.73$ we are still far enough from the 
$K^*$ threshold to safely neglect the $K\pi$ mode with respect
to the single kaon channel.

\subsection{Discussion}

\begin{figure}[t]
   \vspace{-1cm}
   \epsfysize=16cm
   \epsfxsize=14cm
   \centerline{\epsffile{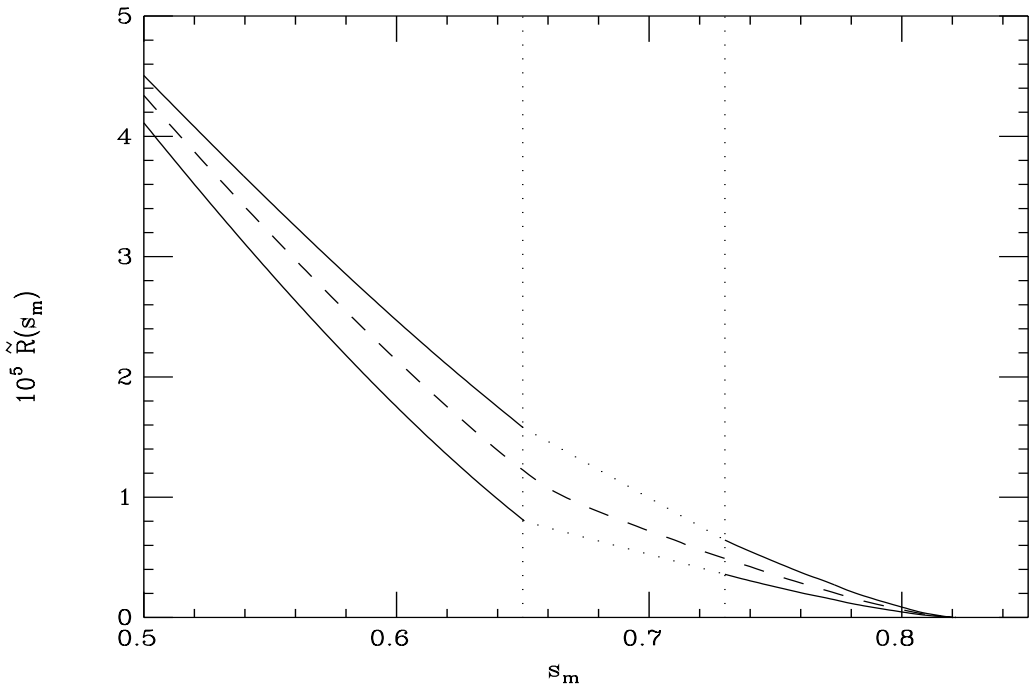}}
   \vspace*{-6cm}
\caption{\label{epspec} The dilepton invariant mass spectrum
$(dB(\bar B\to X_sl^+l^-)/ds_m)/$ $B(\bar B\to X_ce\nu)$ 
$\equiv\tilde R(s_m)$ as a function of $s_m=q^2/M^2_B$.
For $s_m<0.65$ the NLO partonic calculation, including $1/m^2_b$
effects, is used. There the lower, middle and upper curves correspond
to $m_b/\rm{GeV}=4.7$, $4.8$ and $4.9$, respectively. For $s_m>0.73$
we show the HHChPT prediction for $\tilde R(s_m)$, which is dominated
by $\bar B\to\bar Kl^+l^-$. Lower, middle and upper curve are obtained
for $g=0.4$, $0.5$ and $0.6$. Linear interpolations between the
two regions ($0.65< s_m <0.73$) are indicated by dotted lines to guide 
the eye. The dashed curve illustrates a smooth interpolation using
central parameter values. The thresholds for the various exclusive
modes occur at $s_m=0.821$ ($K$), 0.774 ($K\pi$), 0.728 ($K\pi\pi$),
$0.691\pm 0.008$ ($K^*$, $\pm$ half width).} 
\end{figure}

In Fig. \ref{epspec} we compare the HQE result for $R(s)$ with the
HHChPT picture close to the endpoint. 
For this purpose we rescale the
ratio $R(s)$ (\ref{rsnlo}) from quark to physical (hadron) kinematics,
replacing
\begin{equation}\label{rrtilde}
R(s)\to\tilde R(s_m)\equiv 
\frac{M^2_B}{m^2_b} R\left(\frac{M^2_B}{m^2_b} s_m\right)=
\frac{\frac{d}{ds_m}B(\bar B\to X_s l^+l^-)}{B(\bar B\to X_c e\nu)}~.
\end{equation}
This representation of the quark level result is furthermore useful
since it makes the dependence of the prediction on the $b$-quark mass
explicit. The corresponding uncertainty, which will unavoidably exist
in comparing theory with experiment, is illustrated in Fig. \ref{epspec}
for the representative range $m_b=(4.8\pm 0.1)\rm{GeV}$. In this
context we recall that $m_b$ here refers to the pole quark mass.
In fact, since the NLO QCD calculation for $b\to s l^+l^-$ is
available, the distinction of the pole mass from other mass
definitions is already meaningful at first nontrivial (i.e. one-loop)
order. The value of $m_b$ is to be determined from some other observable
and can then be used as input for $\bar B\to X_sl^+l^-$. In principle
the error on $m_b$ can be further reduced in the future.
We remark that the dependence of $\tilde R(s_m)$ on the
renormalization scale $\mu$ ($m_b/2 < \mu < m_b$) is less than
$\pm 5\%$ in the region $0.5<s_m<0.7$.

The $1/m^2_b$ corrections to $\bar B\to X_sl^+l^-$, which are included
in Fig. \ref{epspec}, are negative for $s_m>0.5$,
increase with $s_m$ and reach about
$-20\%$ of the leading result for $s_m=0.65$. As discussed above,
nonperturbative effects that are beyond the control of the HQE
become important for still larger values of $s_m$.

Very close to the endpoint at $s_m=0.821$ HHChPT offers a complementary
approach that may be used to constrain the behaviour of the spectrum
from the region of large $s_m$.
An interpolation suggests itself between the regime $s_m <0.65$,
where the HQE is valid, and $s_m>0.73$, where HHChPT may be used. In
this way an essentially model independent description of the entire
high-$q^2$ region $s_m>0.5$ (above the $\Psi'$ resonance) could be
obtained, at least in principle.
In practice there are however several sizable uncertainties
associated in particular with the HHChPT treatment. The $B^*B\pi$
coupling $g$ is still poorly known. Other uncertainties are related
to the values of $|V_{ts}|$ and $\tau(B_d)$ entering (\ref{rksm}),
but these are less important than the one from $g$. 
Also the $B$ meson decay constant $f_B$ introduces some uncertainty.
\\
A further issue is the reliability of chiral perturbation theory
in the present case. The kaon mass is not very
small with respect to the chiral symmetry breaking scale
$\Lambda_\chi\sim 1.2$~GeV. Thus, even in the vicinity 
of the endpoint, corrections of order 30\%--40\% 
can be expected. In the $K\pi$ channel the situation 
could be even worse, given the presence of the nearby 
$K^*$ resonance. However, for a given value of $s_m$ 
the hadronic invariant mass ranges from 
$M_K$ to $M_{had}^{max}=M_B(1-s_m^{1/2})\simeq 770$~MeV
and only near the upper figure the effect of the 
resonance should be important. 
Given the above remarks, the result for $\tilde R(s_m)$ 
we have presented should still provide a reasonable estimate.
In addition, in view of the kinematical suppression of the $K\pi$ channel,
the fact that the region above $s_m=0.73$ is entirely
determined by $\bar B\to\bar Kl^+l^-$ can be expected to be valid
beyond the limitations of chiral perturbation theory, which is useful
for further studies.

Systematic improvements are possible by going beyond the lowest
order in HHChPT. In \cite{FG} chiral logarithmic corrections to the
leading result have been investigated within HHChPT for the exclusive
mode $\bar B\to\bar Kl^+l^-$. The corrections were found to be about
$40\%$, which is sizable but still moderate enough for the approach
to make sense. 
A related issue is the question of whether to use $f_K$ instead
of $f_\pi$, which also goes beyond the leading order of
chiral perturbation theory.
The calculation of \cite{FG} can be considered as a naive 
estimate of the expected size of the higher-order corrections, but 
lack of knowledge of the corresponding counterterms makes  
any precise statement about their exact value difficult. For this 
reason we have  not explicitly included the chiral logarithms
in our estimates. The related uncertainty is at least partly
included in our variation of the coupling $g$.

Apart from the differential branching fraction also the
forward-back\-ward a\-sym\-me\-try can be studied in HHChPT at large $q^2$.
In this context we note that $A_{FB}$ vanishes identically for the
single kaon mode $\bar B\to\bar Kl^+l^-$. The endpoint of
$A_{FB}$ is therefore determined by $\bar B\to\bar K\pi l^+l^-$
and occurs at $s_m=(1-(M_K+M_\pi)/M_B)^2=0.774$.

We finally remark that the entire high-$q^2$ region (defined 
by $0.5\leq s_m\leq 0.821$) corresponds to an integrated branching
ratio for $\bar B\to X_sl^+l^-$ of about $0.5\cdot 10^{-6}$ in the
Standard Model. Thus, although the dilepton mass spectrum is dropping
to zero towards the endpoint, a sizable branching fraction for
$\bar B\to X_sl^+l^-$ exists in the region that is characterized
by the transition from quark level dynamics to HHChPT. 
The high-$q^2$ regime constitutes one of the interesting regions
to search for $\bar B\to X_sl^+l^-$ in experiment \cite{UA1,CLEO,D0}.
Attempts to describe this part of the spectrum in a model independent
way along the lines proposed in this paper should therefore be useful
for the study of rare $B$ decays at future $B$ physics facilities.

\section{Other nonperturbative corrections}

In addition to higher order terms in the $1/m_b$ expansion, 
$\bar B\to X_sl^+l^-$ decays are affected by long-distance 
corrections related to $c\bar c$ intermediate states. 
These originate from the nonperturbative interactions 
of the $c\bar c$ pair in the process $\bar B\to X_s c\bar c \to X_s 
l^+l^-$. If the dilepton invariant mass is close to one of the 
two narrow $J^{PC}=1^{--}$ $c\bar c$-resonances 
($\Psi(3097)$ and $\Psi'(3686)$) this effect is very 
large and ``obscures'' the short-distance FCNC process. 
However, this background can be eliminated 
by suitable cuts on the dilepton invariant mass. 
Given the vicinity of the two narrow resonances, 
two $q^2$-regions naturally emerge as appropriate  
for the study of short-distance dynamics: the region
below the $\Psi$ and the one above the $\Psi'$. 
In the first case it is  still necessary to deal with the 
$c\bar c$ rescattering below threshold, whereas 
in the second case the effect of broader 
resonances and open charm has to be evaluated.

Nonperturbative contributions generated by $c\bar c$ 
intermediate states have been widely discussed in the literature 
by means of phenomenological resonance-ex\-change 
models \cite{AMM,KS,AH,AHHM}. These approaches are useful near 
the main resonance peaks, but their validity outside this 
region is certainly less reliable. Indeed, the shape of the resonance 
tails far from the peaks is not under control. Moreover, a 
double-counting problem is usually posed by the simultaneous use of 
quark and hadronic degrees of freedom. 
Within this framework, the only way to avoid double counting 
is represented by the approach of \cite{KS} (KS).
Here, in order to take into account charm rescattering,
the correction to $C_9$ induced by $b\to c\bar{c}s$ operators is 
estimated by means of experimental data on 
 $\sigma(e^+e^-\to c\bar c$-hadrons)  using a dispersion relation.
To be more specific, the function $h(z,s)$ appearing in (\ref{c9ep}) 
is replaced by 
\begin{equation}\label{hKS}
h(z,s) \longrightarrow  h(z,0) + {s\over 3} P \int_{s_c}^{\infty}
ds' \frac{ R^{c\bar{c}}_{\rm had} (s') }{s'(s'-s)}
+ i\frac{\pi}{3} R^{c\bar{c}}_{\rm had} (s)~,
\end{equation}
where $R^{c\bar{c}}_{\rm had} (s) = 
\sigma(e^+e^-\to c\bar c)/\sigma(e^+e^-\to \mu^+\mu^-)$ and
$s_c$ is the $c\bar{c}$ threshold.
This method has also the advantage of including open charm contributions.
However, it is exact only in the limit 
where the $\bar B\to X_s c\bar c$ transition can be 
factorized into the product of $\bar{s}b$
and $\bar{c}c$ color-singlet currents 
(i.e. {\it non-factorizable} effects are not included). 
Using this method we have estimated the long-distance 
corrections to the plot in Figure 1. The effect is  
quite small, at the level of several percent, essentially negligible
for $s_m \gsim 0.53$. Below this value the correction 
exceeds 10\% because of the vicinity of the $\Psi'$ peak.

Larger effects from the higher $c\bar c$ resonances
($\Psi(3770)$, $\Psi(4040)$, $\Psi(4160)$, $\Psi(4415)$)
are obtained when a phenomenological factor $\kappa\approx 2.3$
is introduced to enhance resonance production with respect to
the factorization result \cite{LSW}. This is motivated by the fact
that the factorization assumption yields too small values for the
$\bar B\to J/\Psi X_s$ branching fraction. The validity of such a 
procedure for estimating the impact of higher resonances in
$\bar B\to X_sl^+l^-$ is not entirely clear. Further work on
this issue is necessary. In any case the deviations from
quark-hadron duality due to resonances are reduced when the
$\bar B\to X_sl^+l^-$ spectrum is integrated over a large enough range
of $q^2$.

A more systematic and model-independent way to estimate 
$c\bar c$ long-distance effects far from the resonance region,
based on a heavy quark expansion in inverse powers of the 
charm-quark mass, has been recently presented in \cite{BIR}
(see also \cite{CRS}). 
This approach, originally proposed in \cite{VOL} to evaluate 
similar effects in $B\to X_s\gamma$ decays, has the advantage 
of dealing only with partonic degrees of freedom. In this framework 
the leading nonperturbative corrections to $R(s)$ 
turn out to be ${\cal O}(\Lambda^2_{QCD}/m_c^2)$.
They originate from the effective $\bar sb$--photon--gluon vertex
(induced by charm loops), 
where the gluon is soft and couples to the light cloud surrounding 
the $b$ quark inside the $B$ meson. The corresponding 
matrix elements can be related to $\lambda_2$
and thus are known both in magnitude and in sign.
This kind of corrections is complementary to those 
computed in the KS approach, since they are generated by  
the charm rescattering in a color-octet state. Since the 
factorizable corrections vanish for $s\to 0$, as shown by (\ref{hKS}), 
the  ${\cal O}(\Lambda^2_{QCD}/m_c^2)$ effect
is expected to be the dominant long-distance 
contribution for small values of the dilepton invariant mass. 
For $s<0.2$  the relative magnitude of the correction
is very small (at the one or two percent level).
Higher-order terms become more and more important 
near the $c\bar c$ threshold, where the description in terms
of partonic degrees of freedom is clearly inadequate. Using a simple
order-of-magnitude estimate of  higher-order terms, it has been shown 
that the leading corrections should provide a reasonable  
estimate of the effect up to $s=3m_c^2/m_b^2 \approx 0.26$ 
($s_m < 0.21$ ) \cite{BIR}. In this region the 
effect is below 4\%.
The ${\cal O}(\Lambda^2_{QCD}/m_c^2)$
corrections are again very small above the $\Psi'$ peak.

\section{Conclusions}

Within the framework of the heavy quark expansion we have computed
the nonperturbative corrections of ${\cal O}(\Lambda^2_{QCD}/m^2_b)$ to
the dilepton invariant mass spectrum and the lepton forward-backward
asymmetry in $\bar B\to X_sl^+l^-$ decay. Our calculations confirm
the results of \cite{AHHM} for these quantities, which were at
variance with earlier work \cite{FLS}. For completeness we have also
written down the ${\cal O}(\Lambda^2_{QCD}/m^2_b)$ corrections for the
lepton left-right asymmetry.

In the main part of our paper we have then focussed on the region
of high dilepton invariant mass $q^2$ (with $q^2> M^2_{\Psi'}$).
This is one of the relevant search regions in experiments looking
for $\bar B\to X_sl^+l^-$ and corresponds to an integrated branching
ratio of about $0.5\cdot 10^{-6}$ in the Standard Model.
The HQE breaks down for $q^2$ too close to its maximum value at the
endpoint of the dilepton mass spectrum. This is signalled by a
manifest divergence of the relative ${\cal O}(\Lambda^2_{QCD}/m^2_b)$
corrections in the limit $q^2\to m^2_b$, as already observed in 
\cite{AHHM}. We have discussed conceptual aspects of this breakdown of
the HQE for $\bar B\to X_sl^+l^-$ and emphasized that it is impossible
to remedy the failure of the usual $1/m_b$ expansion at the endpoint
by an all-orders resummation, in contrast to the case of e.g. the
photon energy spectrum in $\bar B\to X_s\gamma$.
We were therefore led to consider an alternative, model independent
approach to the endpoint region using HHChPT, which is in principle
well suited in this kinematical regime.
For this purpose we have formulated, at NLO in QCD, an effective
Hamiltonian adapted to the endpoint region. This Hamiltonian is a
variant of the standard Hamiltonian for $b\to sl^+l^-$ transitions
and serves as the basis for calculating the relevant exclusive channels
in the vicinity of $q^2=(M_B-M_K)^2$ within HHChPT.
We explicitly considered the modes $\bar B\to\bar Kl^+l^-$ and
$\bar B\to\bar K\pi l^+l^-$ and demonstrated that the latter is
completely negligible in the kinematical region of interest.
To obtain a complete description of the high-$q^2$ spectrum, an
interpolation between the HHChPT regime and the region of validity
of the heavy quark expansion has been suggested. At present there
are still limitations in accuracy from uncertainties in the
value of $m_b$ and, particularly, in the poorly known HHChPT parameter
$g$ as well as due to neglected higher order terms in the chiral
expansion. However, the approach is essentially model independent
and systematic improvements can in principle be made.

\section*{Acknowledgments}
We thank Stefano Bellucci, Martin Beneke and Gilberto Colangelo
for interesting discussions and a critical reading of the manuscript.

\end{document}